\documentstyle[12pt,fullpage]{article}

\def\half{\textstyle{1 \over 2}}
\def\01{\{0,1\}}
\def\NN{\{0,1,\ldots,2^n\!-\!1\}}
\def\MM{\{0,1,\ldots,2^m\!-\!1\}}
\def\x{\times}

\def\a{\alpha}
\def\b{\beta}

\def\ord{\mbox{ord}_{\,\pi}}
\def\order{\mbox{ord}}

\def\ni{\noindent}

\def\ee{\vspace*{2mm}}
\def\eee{\vspace*{3mm}}
\def\loud#1{\noindent{\bf #1 }}

\newcommand{\qed}{\hfill\rule{2mm}{3mm}}

\begin{document}

\title{The query complexity of order-finding}

\author{Richard Cleve\,%
\thanks{Department of Computer Science, University of Calgary, Calgary, 
Alberta, Canada T2N 1N4.
Email: {\tt cleve@cpsc.ucalgary.ca.}
Partially supported by Canada's NSERC.}
\\ {\small\sl University of Calgary}}

\date{}

\maketitle

\begin{abstract}
We consider the problem where $\pi$ is an unknown permutation on 
$\{0,1,\ldots,2^n-1\}$, $y_0 \in \{0,1,\ldots,2^n-1\}$, and the goal 
is to determine the minimum $r > 0$ such that $\pi^{\,r}(y_0) = y_0$.
Information about $\pi$ is available only via queries that yield 
$\pi^{\,x}(y)$ from any $x \in \{0,1,\ldots,2^m-1\}$ and 
$y \in \{0,1,\ldots,2^n-1\}$ (where $m$ is polynomial in $n$).
The main resource under consideration is the number of these queries.
We show that the number of queries necessary to solve the problem in 
the classical probabilistic bounded-error model is exponential in $n$.
This contrasts sharply with the {\em quantum\/} bounded-error model, 
where a constant number of queries suffices.
\end{abstract}

\section{Introduction}

Let $\pi$ be an arbitrary permutation on $\NN$.
For any $y \in \NN$, define the {\em order of $y$ with respect to $\pi$}, 
denoted as $\ord(y)$, as the minimum $r > 0$ such that 
$\pi^r(y) = y$.
Define $f : \MM \x \NN \rightarrow \MM \x \NN$ as
\begin{equation}
f(x,y) = (x,\pi^{\,x}(y)).
\end{equation}
Note that $f$ can be regarded as a permutation on 
$\01^{m} \x \01^{n} = \01^{m+n}$.

Define the {\em order-finding problem\/} as follows.
As input, one is given $f$ as a black-box.
That is, one can perform queries that return $f(x,y)$ in response 
to $(x,y) \in \{0,1,\ldots,{2^m\!-\!1}\} \x \NN$.
One is also given an element $y_0 \in \NN$.
The goal is to determine $\ord(y_0)$.
The resource under consideration is the number of queries 
performed.

Shor's remarkable algorithm for integer factorization on a quantum 
computer \cite{Sho94} is based on solving the {\em modular order-finding 
problem}.
In this problem, the input is an $n$-bit integer $N$ and also an integer 
$a$ such that $0 < a < N$ and $\gcd(a,N) = 1$.
The goal is to find the minimum $r > 0$ such that $a^r \bmod N = 1$.
This is equivalent to a specialized instance of the order-finding problem 
defined above with $y_0 = 1$, and 
\begin{equation}
\pi(y) = \cases{ (a y) \bmod b & if $0 \le y < N$ \cr
                 y             & if $N \le y < 2^n$. \cr}
\end{equation}
The quantum algorithm in \cite{Sho94} actually solves the more general 
order-finding problem with $m = 2n$, and it accomplishes this with only 
{\em two\/} queries and $O(n^2)$ auxiliary operations (measured in terms 
of, say, two-qubit quantum gates).

We investigate the {\em classical\/} query complexity of the general 
order-finding problem, and our main results are the following.\ee

\loud{Theorem 1:}{\sl Any classical deterministic procedure 
for the order-finding problem requires $\Omega(\sqrt{2^n \over m})$ 
queries (assuming $m \ge n$).}\eee

\loud{Theorem 2:}{\sl Any classical probabilistic procedure for 
the order-finding problem requires $\Omega({2^{n/3} \over \sqrt{m}})$ 
queries if the success probability is bounded above zero 
(assuming $m \ge n$).}\ee

In particular, when $m = 2n$, the quantum vs.\ classical query 
complexity is $O(1)$ vs.\ $\Omega({2^{n/3} \over \sqrt{n}})$ in the 
bounded-error model.
A comparison with other known quantum vs.\ classical query separations 
in the bounded-error model is given in Table~1.

\begin{table}[h]
\centering
\parbox{135mm}{
\caption{Comparison of quantum vs.\ classical separations for query problems 
in the bounded-error model.}\vspace*{1mm}}
\begin{tabular}{|l|c|c|c|}
\hline
References & number of bits & quantum upper bound 
& classical lower bound \\
\hline
\hline
{\small Bernstein \& Vazirani} \cite{BV93} 
& $n\!+\!1$ & $O(1)$ & $\Omega(n)$ \\
\hline 
{\small Bernstein \& Vazirani} \cite{BV93} 
& $\Theta(n)$ & $n^{O(1)}$ & $n^{\Omega(\log n)}$ \\
\hline 
{\small Simon} \cite{Sim94} 
& $2n$  & $O(n)$ & $\Omega(2^{n/2})$ \\
\hline 
{\small Grover} \cite{Gro96} 
& $n\!+\!1$ & $O(2^{n/2})$ & $\Omega(2^n)$ \\
\hline
{\small Shor} \cite{Sho94} / present result 
& $3n$ & $O(1)$ & 
$\Omega(2^{n/3}/\sqrt{n}\,)$ \\
\hline
\end{tabular}
\end{table}

Our classical lower bounds for order-finding are exponential whenever 
$m$ is polynomial in $n$ (and even for some settings of $m$ that are 
exponentially larger than $n$, such as $m = 2^{n/2}$).

It is sometimes stated informally that the ``period-finding'' task 
performed by the quantum Fourier transform in Shor's algorithm 
\cite{Sho94} cannot be accomplished efficiently by any classical method.
Theorem~2 can be viewed as a confirmation of this in a formal setting.%
\footnote{In the context of the {\em modular\/} order-finding problem, 
no interesting classical lower bound is known, and such a lower bound 
would constitute a major breakthrough in computational complexity theory.}

It should be noted that classical order-finding methods that are not 
entirely trivial exist, since it can be advantageous to perform 
queries that request $\pi^{\,x}(y)$ where $x$ is much larger than $2^n$.
For example, consider the case where $n=4$ and $m=7$, so the potential 
values of $\ord(y_0)$ are $\{1,2,\ldots,16\}$.
We first state the following lemma, which is simple to prove.\ee

\loud{Lemma 3:}{\sl $\pi^{\,x}(y) = y$ if and only if $\ord(y)\,|\,x$.}\ee

\ni Now, after a single query requesting $\pi^{90}(y_0)$ is performed, 
the possible values of $\ord(y_0)$ are reduced by a 
factor of two: if $\pi^{90}(y_0) = y_0$ then 
$\ord(y_0) \in \{1,2,3,5,6,9,10,15\}$; otherwise, 
$\ord(y_0) \in \{4,7,8,11,12,13,14,16\}$.
This process can be continued.
For example, suppose that $\pi^{90}(y_0) \neq y_0$.
Then let the second query request $\pi^{56}(y_0)$.
If $\pi^{56}(y_0) = y_0$ then $\ord(y_0) \in \{4,7,8,14\}$; otherwise, 
$\ord(y_0) \in \{11,12,13,16\}$.
It is straightforward to extend this to an algorithm that, for these 
settings of $n$ and $m$, always deduces $\ord(y_0)$ with four queries.

Theorems 1 and 2 imply, among other things, that the binary splitting 
which occurs in the above example cannot occur for larger values of $n$.
Informally, the basic idea behind the proofs is that there are many 
potential values of $\ord(y_0)$ which are large primes, and an $x \in \MM$ 
cannot have too many of these as divisors.
Thus, on average, a query of the form $\pi^{\,x}(y)$ eliminates very few 
of these values.
The technicalities in the proofs arise from considering the ways that 
information can accumulate from a sequence of several queries.

Formally, the procedures that we are analyzing are {\em decision trees}, 
which have a query at each internal node, and a child node corresponding 
to each possible outcome of that query.
Each leaf has an output value associated with it.
The execution of a decision tree is a path from the root to a leaf 
that follows the outcomes of the queries.
The depth of the tree corresponds to the number of queries of the procedure 
(for a worst-case input).
A {\em randomized decision tree\/} represents a decision precedure that 
is allowed to flip coins and have its behavior depend on the outcomes.
It can be defined formally as a probability distribution on a set of 
deterministic decision trees.

\section{Lower bound for deterministic decision trees}

In this section, we prove Theorem~1.
The proof is based on the {\em evasive\/} method.
Let the query algorithm (decision tree) be fixed and construct 
a sequence of responses to queries which are consistent with 
at least two permutations $\pi_1$ and $\pi_2$ such that 
$\order_{\pi_1}(y_0) \neq \order_{\pi_2}(y_0)$.
Then the length of this sequence is a lower bound on the query 
complexity of the problem.

Define the set 
\begin{equation}
R = \{\, r : \mbox{$r$ is prime and $2^{n-1} < r \le 2^n$}\}.
\end{equation}
We will consider the restricted set of permutations, for which 
$\order_{\pi}(y_0) \in R$.
This is not a very severe restriction because, by the Prime Number Theorem 
(see, for example, \cite{BS96}), the following is a lower bound on the 
size of $R$.\ee

\loud{Lemma 4:}{\sl The size of $R$ is at least $\a {2^n \over n}$, 
where $\a = 0.721$ (for sufficiently large $n$).}\ee

\ni Intuitively, the next lemma asserts that, since the elements of $R$ 
are primes of significant size, the number that are eliminated by a 
query is not very large.\ee

\loud{Lemma 5:}{\sl For any $x < 2^h$ the number of elements of $R$ 
that divide $x$ is at most ${h \over n-1}$.}\ee

\loud{Proof:}
If $x$ contains more than ${h \over n-1}$ divisors from $R$ then 
$x > (2^{n-1})^{h \over n-1} = 2^h$, a contradiction.\qed\ee

Now, to construct the evasive sequence of responses, it is helpful to have a 
systematic way of keeping track of the evolution of information 
about the unknown permutation $\pi$ that unfolds as the queries 
occur.
Define a {\em chain\/} as a weighted linked-list of the form illustrated 
in Figure 1, where $k \le 2^n$, $y_1,y_2,\ldots,y_k$ are {\em distinct\/} 
elements of $\NN$, and $w_1,\ldots,w_{k-1} \in \MM$.\
\setlength{\unitlength}{1mm}
\begin{figure}[h]
\centering
\begin{picture}(130,20)(0,0)

\put(10,10){\circle{10}}
\put(40,10){\circle{10}}
\put(70,10){\circle{10}}
\put(120,10){\circle{10}}

\put(15,10){\vector(1,0){20}}
\put(45,10){\vector(1,0){20}}
\put(75,10){\vector(1,0){15}}
\put(100,10){\vector(1,0){15}}
\put(95,10){\makebox(0,0){$\cdots$}}

\put(10,10){\makebox(0,0){$y_1$}}
\put(40,10){\makebox(0,0){$y_2$}}
\put(70,10){\makebox(0,0){$y_3$}}
\put(120,10){\makebox(0,0){$y_k$}}

\put(25,12){\makebox(0,0){$w_1$}}
\put(55,12){\makebox(0,0){$w_2$}}
\put(82,12){\makebox(0,0){$w_3$}}
\put(107,12){\makebox(0,0){$w_{k-1}$}}

\end{picture}
\parbox{135mm}{\caption{A chain of length $k$.}}
\end{figure}
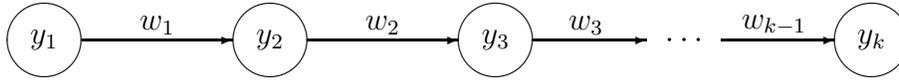
A link with weight $w_i$ from $y_i$ to $y_{i+1}$ indicates that 
$\pi^{\,w_i}(y_i) = y_{i+1}$.
Several other relationships follow by transitivity: 
$\pi^{\,w_i + \cdots + w_{j-1}}(y_i) = y_j$, for each 
$i, j \in \{1,2,\ldots,k\}$ with $i < j$.
After each query is made and responded to, the chain is adjusted 
so as to contain all properties of $\pi$ that have been determined 
up to that point in the execution of the query algorithm.

Call a query {\em internal\/} if it requests $\pi^{\,x}(y)$, where 
$y \in \{y_1,\ldots,y_k\}$, or if it is the very first query.
There are two possibilities with an internal query.
One is that all the information about the response is already contained 
in the existing chain, in which case this information is simply returned 
and the chain does not need to be adjusted.
The second possibility is that the information is not yet determined by 
the existing chain.
An example is the query requesting $\pi^{\,x}(y_1)$, where 
$w_1 < x < w_1 + w_2$.
In this case, the information returned is some (arbitrary) 
$y \not\in \{y_1,\ldots,y_k\}$ and the chain is updated to reflect 
this.
For the given example, the updated chain would contain a new element 
between element $y_2$ and $y_3$.
Note that the property that the weights are all in 
$\MM$ is preserved.
We will also have to consider {\em external\/} (i.e.\ non-internal) 
queries, requesting $\pi^{\,x}(y)$, where $y \not\in \{y_1,\ldots,y_k\}$, 
but we postpone this until later.

Suppose that, after a number of queries, the resulting chain is 
that of Figure 1.
Thus, $\pi$ can be any permutation consistent with this chain.
The elements of the chain must all be in the same cycle of $\pi$.
What are the possible sizes of this cycle?\ee

\loud{Lemma 6:}{\sl For any $r \in R$, the chain of 
Figure 1 is consistent with cycle size $r$ if and only if 
$r \!\not|\,\, w_i + \cdots + w_{j-1}$ for all $i, j \in \{1,2,\ldots,k\}$ 
with $i < j$.}\ee

\loud{Proof:}
For the ``only if'' direction, if $r \,|\, w_i + \cdots + w_{j-1}$ 
then, by Lemma~3, $y_i = y_j$, which contradicts the fact that $y_i$ 
is distinct from $y_j$.
For the ``if'' direction, suppose that 
$r \!\not|\,\, w_i + \cdots + w_{j-1}$ (for all $i<j$) and map the chain 
onto a cycle of size $r$.
Then, for all $i < j$, $y_i$ will not collide with $y_j$, since, by Lemma~3, 
this would imply that $r \,|\, w_i + \cdots + w_{j-1}$.
\qed\ee

\ni Let us now consider how many cycle sizes $r \in R$ are consistent with 
the chain of Figure~1.
There are ${k(k-1) \over 2} < \half k^2$ values of $i, j \in \{1,2,\ldots,k\}$ 
with $i < j$.
For each such pair, $w_i + w_{i+1} + \cdots + w_{j-1} < k2^m \le 2^{n+m}$, 
so, by Lemma 5, the number of its divisors that reside in $R$ is at most 
$m+n \over n-1$.
Therefore, by Lemma~4, at least 
$\a {2^n \over n} - \half k^2 ({m+n \over n-1})$ 
different values in $R$ are consistent with the chain of Figure~1.
It follows that $r$ is not uniquely determined until 
$\a {2^n \over n} - \half k^2 ({m+n \over n-1}) < 2$ 
which means 
\begin{eqnarray}
k & > & 
\sqrt{2
{\textstyle{
\left({n-1 \over n}\right)\left({\a 2^n - 2n \over m+n}\right)}
}} 
\ \in \ \Omega\left(\sqrt{2^n / m}\right).
\end{eqnarray}
We now address the case of external queries.
For an external query requesting $\pi^{\,x}(y)$, where 
$y \not\in \{y_1,\ldots,y_k\}$, $y$ might not be in the cycle 
containing the elements of the existing chain.
Or $y$ might be in this cycle, but at an unspecified place.
This information could be recorded by starting a new chain, and 
the resulting data structure after several queries might consist 
of several chains.
To simplify the evasive procedure, the following two steps are 
performed.
First, a new element $y$ is added to the beginning of the chain with a 
weight of 1.
Then the procedure for an internal query is followed.
Note that the resulting chain actually specifies more information about 
$\pi$ than revealed by the queries (since the queries do not reveal 
that $\pi^{1}(y) = y_1$).
This is not a problem because what we are using is the fact that the chain 
contains {\em at least\/} as much information about $\pi$ as the queries 
have revealed.
After $k$ (internal or external) queries, the result is a single 
chain of length at most $2k$.
It follows that an evasive sequence of length 
$\Omega(\sqrt{2^n/m})$ exists, completing the proof of 
Theorem~1.

\section{Lower bound for randomized decision trees}

To prove Theorem~2, we use the game theoretic approach of Yao \cite{Yao83}, 
and exhibit a probability distribution on the set of permutations on 
$\NN$ for which every {\em deterministic\/} decision tree must make 
$\Omega({2^{n/3} \over \sqrt{m}})$ queries in order to determine $r$ with 
probability at least $2 \over 3$ (say).
It then follows that, for any {\em randomized\/} decision tree 
(which corresponds to a probability distribution on deterministic 
decision trees), $\Omega({2^{n/3} \over \sqrt{m}})$ queries are necessary 
to determine $r$ with probability at least $2 \over 3$.

Define a {\em collision\/} as any query requesting $\pi^{\,x}(y)$ 
with $x > 0$ whose response is $y$ (i.e.\ $\pi^{\,x}(y) = y$).
It suffices to show that $\Omega({2^{n/3} \over \sqrt{m}})$ queries are 
necessary to obtain a collision with probability at least $2 \over 3$.
This is because any execution of a decision tree that correctly 
determines $r$ can be adjusted to include a collision with at most 
one additional query (requesting $\pi^{\,r}(y_0)$).

Assign a probability distribution to the set of permutations on $\NN$ 
as follows.
First (assuming for convenience that $n$ is divisible by 3), choose an 
order $r$ uniformly from the set 
\begin{equation}
R^{\prime} = 
\{\, r : \mbox{where $r$ is prime and $2^n-2^{2n/3} < r \le 2^n$}\}.
\end{equation}
Estimating the size of $R^{\prime}$ is more subtle than for $R$; 
however, sufficient lower bounds do exist (the relevant result is 
implicit in \cite{Mon69}, explicitly stated in \cite{IJ79}, and the 
value value of $\b$ in the lemma below is from \cite{BH96}).\ee

\loud{Lemma 7 \cite{Mon69,IJ79,BH96}:}{\sl The size of $R^{\prime}$ 
is at least $\b {2^{2n/3} \over n}$, where $\b = {1 \over 14}$ 
(for sufficiently large $n$).}\ee

\ni Once $r$ is chosen, the generation of $\pi$ proceeds as follows.
Let $\pi$ consist of two cycles, one of size $r$ and one of size 
$s = 2^n\!-\!r$.
The $r$-cycle consists of $r$ randomly selected elements of $\NN$ inserted 
in a random order, and the $s$-cycle consists of the remaining 
$s$ elements of $\NN$ inserted in a random order.
With probability at least $1-2^{-n/3}$, $y_0$ is in the $r$-cycle.
The permutation $\pi$ can be explicitly represented by an array 
$A = (A_0,A_1,\ldots,A_{2^n-1})$ and the value $r$ with the understanding 
that $s = 2^n\!-\!r$ and 
\begin{equation}
\label{array}
\pi^{\,x}(A_i) = \cases{A_{(i+x) \bmod r}           & if $0 \le i < r$\cr
                        A_{((i-r+x) \bmod s) + r} & if $r \le i < 2^n$.\cr}
\end{equation}

To construct $\pi$, one could choose $r$ as above and then insert the 
values of $\NN$ into $A$ in a random order.
To simulate the execution of any fixed decision tree $T$, the responses 
to queries can be made by referring to $A$; however, we describe an 
alternate way of responding to the queries in $T$ which is stochastically 
equivalent to this.
In the alternate method, the entries of $A$ are determined ``on the fly'', 
as the queries are received.
To begin with, three items are randomly created:
\begin{itemize}
\item A list $V$ of ``new values'', $v_0, v_1, \ldots, v_{2^n-1}$ 
(the elements of $\NN$ in a random order).
An {\em access\/} to this list returns the first item, and then 
removes this item from the list (so the next access returns the 
second item, and so on).
\item A list $I$ of ``new indices'', $i_0, i_1, \ldots, i_{2^n-1}$ 
(the elements of $\NN$ in a random order).
An {\em access\/} to this list returns the first item, and then 
removes this item from the list.
\item A random $r \in R^{\prime}$.
\end{itemize}
The array $A$ is initially empty.
Then, whenever a query requesting $\pi^{\,x}(y)$ is made, the following 
two-stage procedure is carried out to update $A$.
\begin{enumerate}
\item 
The value of $i$ such that $A_i = y$ is determined.
If $y$ has not yet been inserted into $A$, then it is inserted in the 
following way.
The elements of $I$ are accessed until one occurs that corresponds to 
an $i$ such that $A_i$ has not yet been assigned a value.
Then $A_i$ is assigned the value $y$.
\item
The value of the $j$ corresponding to $A_j = \pi^{\,x}(A_i)$ (according 
to Eq.~\ref{array}) is determined.
Then, if $A_j$ has not yet been assigned a value, the elements of 
$V$ are accessed until a value that has not yet appeared in $A$ 
occurs, and $A_j$ is assigned to that value.
\end{enumerate}
Finally, the value of $A_j$ is the response to the query.

The decision tree $T$ contains $N$ branches from every query.
However, once $V$ has been determined (but independent of $I$ and $s$), 
there is always at most one branch possible that corresponds to a 
``new value'' from $V$ (i.e.\ where the query results in accesses to $V$ 
in Step~2).
For example, suppose that the very first query is $(x,y)$.
Then one possible branch is $y$ (if $\pi^{\,x}(y) = y$), and the only 
other possible branch is $v^{\prime}$ (if $\pi^{\,x}(y) \neq y$), 
where $v^{\prime}$ is a value accessed from $V$ (specifically, 
$v^{\prime} = v_0$ if $v_0 \neq y$; and $v^{\prime} = v_1$ if $v_0 = y$).
The latter branch corresponds to a ``new value''.
We shall consider the path from the root to a leaf that follows 
the new value branch whenever such a branch is possible (if a new value 
branch is not possible then the value of the query is determined 
by the previous queries, so only one branch is possible, and that 
is the one taken in this path).
Call this path the {\em principal path\/} of $T$.

We now describe a procedure for associating a chain with every 
query along the principal path of $T$.
The chain associated with each query subsumes all the information 
about $\pi$ that would be determined up to and including that query, 
if the principal path were taken up to that point.
These chains depend on $I$ (as well as $V$, which determines the 
principal path) and may fail with a certain probability (that we will 
show to be negligibly small).
For the first query requesting $\pi^{\,x}(y)$, if the first new address 
$i_0$ does not exceed $2^n - 2^{2n/3}$, we assign the chain of length 
two of Figure~2; otherwise the process fails.
\setlength{\unitlength}{1mm}
\begin{figure}[h]
\centering
\begin{picture}(50,20)(0,0)

\put(10,10){\circle{10}}
\put(40,10){\circle{10}}

\put(15,10){\vector(1,0){20}}

\put(10,10){\makebox(0,0){$y$}}
\put(40,10){\makebox(0,0){$v^{\prime}$}}

\put(25,12){\makebox(0,0){$x$}}

\end{picture}
\parbox{135mm}{\caption{The chain associated with the first query 
of the principal path.}}
\end{figure}
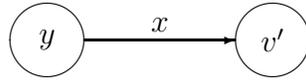
This corresponds to $\pi^{\,x}(y) = v^{\prime}$.
Note that the head of the chain ($y$) is in a definite position ($i_0$) 
in array $A$, determined by $V$ and $I$, but independent of the 
value of $r$.
We call $i_0$ the {\em location of the head of the chain}.
Also, note that, since $i_0 \le 2^n - 2^{2n/3} < r$, both $y$ and 
$v^{\prime}$ are in the $r$-cycle of $\pi$ (whatever the value of $r$ is).

For each subsequent query in the principal path, the chain is updated 
to include the information revealed by this query in the following way.
Assume that the chain associated with the previous query is of the form 
in Figure 1 and that $i^{\prime}$ is the location of the head of the 
chain ($y_1$).
We consider the case of internal and external queries separately.
For {\em internal\/} queries, the chain is updated in the natural way, 
as in the proof of Theorem~1, with the value of a possible new node 
taken from $V$.
The location of the head of the chain remains $i^{\prime}$.

The procedure for {\em external\/} queries is a little more 
complicated.
First, let $i^{\prime\prime}$ be the next element of $I$.
If $i^{\prime\prime}$ exceeds $2^n - 2^{2n/3}$ then the procedure fails.
Otherwise, a new node is inserted into the chain at a place 
dependent on the value of $i^{\prime} - i^{\prime\prime}$.
If $i^{\prime} - i^{\prime\prime} > 0$ then the new node is linked 
before the head of the chain with a link of weight 
$i^{\prime} - i^{\prime\prime}$, as illustrated in Figure~3, and the 
location of the head of the chain is changed to $i^{\prime\prime}$.\
\setlength{\unitlength}{1mm}
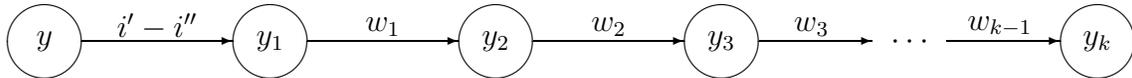
\begin{figure}[h]
\centering
\begin{picture}(160,20)(0,0)

\put(10,10){\circle{10}}
\put(40,10){\circle{10}}
\put(70,10){\circle{10}}
\put(100,10){\circle{10}}
\put(150,10){\circle{10}}

\put(15,10){\vector(1,0){20}}
\put(45,10){\vector(1,0){20}}
\put(75,10){\vector(1,0){20}}
\put(105,10){\vector(1,0){15}}
\put(130,10){\vector(1,0){15}}
\put(125,10){\makebox(0,0){$\cdots$}}

\put(10,10){\makebox(0,0){$y$}}
\put(40,10){\makebox(0,0){$y_1$}}
\put(70,10){\makebox(0,0){$y_2$}}
\put(100,10){\makebox(0,0){$y_3$}}
\put(150,10){\makebox(0,0){$y_k$}}

\put(25,12){\makebox(0,0){$i^{\prime} - i^{\prime\prime}$}}
\put(55,12){\makebox(0,0){$w_1$}}
\put(85,12){\makebox(0,0){$w_2$}}
\put(112,12){\makebox(0,0){$w_3$}}
\put(137,12){\makebox(0,0){$w_{k-1}$}}

\end{picture}
\parbox{135mm}{
\caption{First step in updating the chain for an external query requesting 
$\pi^{\,x}(y)$ when $i^{\prime} - i^{\prime\prime} > 0$.}}
\end{figure}
If $i^{\prime} - i^{\prime\prime} < 0$ then the new node is linked 
after the head of the chain, in an appropriate position so as to have 
weighted distance $i^{\prime\prime} - i^{\prime}$ from the head.
It is possible that this causes an ``overlap'' in that there is already 
a node in the chain with weighted distance $i^{\prime\prime} - i^{\prime}$ 
from the head.
In this event, the process fails.
After the node has been inserted into the chain, the query is processed 
exactly as an internal query.

The procedure of associating chains with queries continues until either 
the end of the principal path is reached or a failure occurs.
If $t$ is the depth of $T$ then the probability of termination due to 
failure is bounded above by $t 2^{-n/3} + t^2 (2^n - 2^{2n/3})^{-1}$ 
(which is $o(1)$ if $t \in o(2^{n/3})$).

To recap so far, based on $V$ and $I$ (but independent of the choice of $r$), 
a principal path from the root until a leaf of decision tree $T$ 
is determined (with a negligible failure probability $o(1)$).
Consider the ``final'' chain, associated with the last query along the 
principal path.
This chain has length $k \le 2t$, and it is completely independent 
of the choice of $r$.
Moreover, since this chain subsumes all the information obtained about 
the permutation $\pi$, no collision occurs whenever an execution of 
$T$ follows the principal path.

Now, consider the probability (with respect to the random choice 
of $r \in R^{\prime}$) of the event that the principal path is not taken 
(assuming that the final chain has length $k$).
By Lemma~6, this event occurs whenever 
$r \!\not|\,\, w_i + \cdots + w_{j-1}$ 
for all $i, j \in \{1,2,\ldots,k\}$ with $i < j$.
The probability of this is bounded below by 
\begin{equation}
{\b {2^{2n/3} \over n} - \half k^2 ({m+n \over n - 1}) \over 
\b {2^{2n/3} \over n}} 
\ \ = \ \ 1 \ - \ {k^2 \over 2} \left({m+n \over \b 2^{2n/3}} \right) 
\left({n \over n-1}\right),
\end{equation}
which is bounded above $1 \over 3$ unless 
\begin{equation}
k \ \ge \ \sqrt{{4 \over 3}\left({n-1 \over n}\right) 
\left({\b 2^{2n/3} \over m+n}\right)} 
\ \in \ \Omega\left({2^{n/3} \over \sqrt{m}}\right).
\end{equation}
From this, Theorem~2 follows.

\section{Upper bounds}

When $m \ge n+1$, there is a probabilistic procedure that solves the 
order-finding problem with $O(\sqrt{2^n})$ queries.
The idea is to select $x_1,x_2,\ldots,x_k \in \{0,1,\ldots,2^{n+1}\!-\!1\}$ 
randomly, and output the minimum positive $x_i - x_j$, where 
$i, j \in \{1,2,\ldots,k\}$ and $\pi^{\,x_i}(y_0) = \pi^{\,x_j}(y_0)$.
The probability that that the output is {\em not\/} $\ord(y_0)$ is 
bounded above by $2^{-O(k^2/2^n)}$.
There is a setting $k \in O(\sqrt{2^n})$ that bounds this below any 
positive constant.

\section*{Acknowledgments}
Guidance from Richard Mollin and John Watrous with number theoretic 
literature is gratefully acknowledged.

\end{document}